\documentstyle[12pt]{article}
\begin{document}
\begin{titlepage}
\baselineskip=0.3in
\begin{flushright}
HNU-ITP-96-1
\end{flushright}
\vspace{0.3in}

\begin{center}
{\large Top-squark mixing effects in the supersymmetric \\
electroweak corrections to top quark production at the Tevatron}
\vspace{.5in}

 Jin Min Yang$^{a,b}$ and Chong Sheng Li$^{a,c}$ 
\vspace{.5in}

$^a$CCAST(World Laboratory)\\
P.O.Box 8730, Beijing 100080, P. R. China \\
\bigskip$^b$ Physics Department, Henan Normal University,\\
Xinxiang, Henan 453002, P.R.China \footnote{mailing address}\\
\bigskip$^c$ Physics Department, PeKing University\\
Beijing 100871, P. R. China
\end{center}
\vspace{.6in}

\begin{footnotesize}
\begin{center}\begin{minipage}{5in}

\begin{center} ABSTRACT\end{center}

Taking into account the mixing effects between left- and right-handed 
top-squarks, we calculate the genuine supersymmetric eletroweak correction
to top quark production at the Tevatron in the minimal supersymmetric model. 
The analytic expressions of the corrections to both the parton level cross 
section and the total hadronic cross section are presented. Some numerical 
examples are also given to show the size of the corrections.  

\end{minipage}\end{center}
\end{footnotesize}
\vspace{.8in}

PACS number: 14.80Dq; 12.38Bx; 14.80.Gt
\end{titlepage}
\eject
\baselineskip=0.3in

\begin{center}1. Introduction \end{center}

 The top quark has been discovered by CDF and D0 collaborations at the 
Tevatron $[1]$. The mass and production cross section are found to be 
$176\pm 8(stat)\pm 10(syst) \left (199^{+19}_{-21}(stat)\pm 22(syst)\right )$
GeV and $6.8^{+3.6}_{-2.4} \left ( 6.4\pm 2.2\right ) Pb$ by CDF ( DO ) 
collaboration. The comparison of the theoretical calculation of the top quark 
production cross section with experimental results is necessary in the test 
of the mechanism by which top quarks are produced. Within the framework of 
the Standard Model (SM) the next-to-leading-order calculation for the QCD
processes was completed several years ago$[2]$. Recent works $[3]$ have 
extended those results with the inclusion of the exact order $\alpha_s^3$
corrected cross section and the resummation of the leading soft gluon 
corrections in all orders of perturbation theory. The cross section was 
predicted to be $\sigma_{t\bar t}(m_t=176 GeV)=4.79^{+0.67}_{-0.41} Pb[3]$. 
The latest results given by Berger $[4]$ was $\sigma_{t\bar t}(m_t=175 GeV)
=5.52^{+0.07}_{-0.45} Pb$. The one-loop electroweak corrections to the cross 
section were found to be only a few percent$[5]$. Therefore, the results of 
theoretical prediction in the SM are almost consistent with the experimental 
results within the region of the errors.

Since the corrections to top quark production cross section above 20\% 
are potentially observable at the Tevatron, it is tempting to calculate 
the radiative corrections arising from the new physics beyond the SM.
In the minimal supersymmetric model (MSSM), the Yukawa correction and 
supersymmetric QCD correction were calculated in Refs.[6,7]. These 
corrections  cannot reach the observable level for experimentaly 
allowed parameter values. The genuine supersymmetric electroweak 
corrections  of order  $\alpha m_t^2/m_W^2$, which arise from loops of 
chargino, neutralino and squark, have also been calculated by us in Ref.[8] 
and its erratum [9], where we neglected the mixings between left- and 
right-handed squarks and assumed the mass degeneracy for all squarks. In 
such a simple case, the analytic results were quite simple and the numerical 
size of the corrections could not reach the observable level for squark mass 
heavier than 100 GeV. However, due to the possible significant mixing effects 
between left- and right-handed top-squarks, which is suggested by low-energy 
supergravity models but is completely general [10], the mass splitting between 
the two mass eigenstates of top squark may be quite large. The supersymmetric 
electroweak corrections may be sensitive to top-squark mixing effects.

In this paper, taking into account the mixing effects between left- and 
right-handed top-squarks, we present the genuine supersymmetric eletroweak 
correction to top quark production at the Tevatron in the minimal 
supersymmetric model. In Sec.2, we briefly overview top-squark mixing.
In Sec.3, we give the analytic expressions of the corrections to both 
the parton level cross section and the total hadronic cross section.
In Sec.4 we present some numerical examples to show the size of the 
corrections.  

\begin{center} 2. Top-squark mixing\end{center}

  The mass matrix of top-squarks takes the form [10]
\begin{eqnarray}
-L_m& =& (\tilde t^*_L \ \tilde t^*_R)\left ( \begin{array}{cc}
m_{\tilde t_L}^2 & m_t M_{LR}\\m_t M_{LR} & m_{\tilde t_R}^2 
\end{array}
\right )\left ( \begin{array}{ll} \tilde t_L \\ \tilde t_R
               \end{array}\right )\nonumber\\
m_{\tilde t_L}^2& =& M^2_{\tilde t_L}+m_t^2+\cos(2\beta)(\frac{1}{2}
-\frac{2}{3}\sin^2 \theta_W)M_Z^2\nonumber\\
m_{\tilde t_R}^2& =& M^2_{\tilde t_R}+m_t^2
               +\frac{2}{3}\cos(2\beta)\sin^2 \theta_W M_Z^2\nonumber\\
M_{LR}&=&\mu\cot\beta+A_t \tilde M
\end{eqnarray}
where $ M^2_{\tilde t_L}, M^2_{\tilde t_R}$ are the soft SUSY-breaking mass
terms for left- and right-handed top-squarks,$\mu$ is the coefficient of
the $H_1H_2$ mixing term in the superpotential, $A_t\tilde M$ is the 
coefficient of the dimension-three trilinear soft SUSY-breaking term
$\tilde t_L\tilde t_R H_2$, and $\tan\beta=v_2/v_1$ is the ratio of the 
vacuum expectation values of the two Higgs doublets.  

The mass eigenstates of top-squark are obtained by  
\begin{eqnarray}
\left ( \begin{array}{l} \tilde t_1\\ \tilde t_2 \end{array} \right )
=R \left ( \begin{array}{ll} \tilde t_L\\ \tilde t_R \end{array} \right )
=\left ( \begin{array}{ll} \cos\theta & \sin\theta\\ 
           -\sin\theta & \cos\theta \end{array} \right ) 
\left ( \begin{array}{ll} \tilde t_L\\ \tilde t_R \end{array} \right )
\end{eqnarray}
and the masses of $\tilde t_{1,2}$ are given by 
\begin{equation}
R~~ M^2_{\tilde t}~~ R^T=\left ( \begin{array}{ll} m^2_{\tilde t_1}& 0\\ 
		0 & m^2_{\tilde t_2} \end{array} \right )
\end{equation}
The expressions of $\theta$ and $m^2_{\tilde t_{1,2}}$ are given by
\begin{eqnarray}
\tan 2\theta&=&\frac{2M_{LR}m_t}{m_{\tilde t_L}^2-m_{\tilde t_R}^2} \\
m^2_{\tilde t_{1,2}}&=&\frac{1}{2}\left [ m_{\tilde t_L}^2
          +m_{\tilde t_R}^2\mp\sqrt{(m_{\tilde t_L}^2-m_{\tilde t_R}^2)^2
          +4M_{LR}^2m_t^2 } \right]
\end{eqnarray}

For sbottoms, since we neglect the mixing between left- and right-handed
sbottoms, we have
\begin{equation}
m^2_{\tilde b_{1,2}}=m^2_{\tilde b_{L,R}}=m^2_b+M^2_{\tilde b_{L,R}}
	\pm \cos(2\beta)(T^3_{L,R}-Q_b\sin^2 \theta_W)M_Z^2
\end{equation}
where $T^3_{L,R}=-\frac{1}{2},0$ and $Q_b=-\frac{1}{3}$.  $M_{\tilde b_{L,R}}$ 
are the soft SUSY-breaking mass terms for left- and right-handed sbottoms. 

\begin{center} 3.  Analytical expression of the correction\end{center}

    At the Tevatron, the top quark is dominantly produced via quark-antiquark
annihilation $[11]$. The genuine supersymmetric electroweak correction
of order $\alpha m_t^2/m_W^2$ to the amplitude is contained in the correction
to the vertex of top-quark color current. The  relavent Feynman diagrams are 
shown in Fig.1 in Ref.[7]. The  Feynman rules can be found in Ref.[12]. In 
our calculation, we use dimensional regularization to regulate all the 
ultraviolet divergences in the virtual loop corrections and we adopt the 
on-mass-shell renormalization scheme[13]. The renormalized amplitude for 
$q\bar q \rightarrow t \bar t$ can be written as
\begin{equation}
{\it M_{ren}}= {\it M_0}+\delta {\it M},
\end{equation}
where ${\it M_0}$ is the amplitude at tree-level and $\delta {\it M}$
is the correction to the amplitude, which are given by
\begin{eqnarray}
{\it M_0}&=&\bar {\it v}(p_2)(-ig_sT^A \gamma^{\nu}){\it u}(p_1)
	~\frac {-i g_{\mu\nu}} {\hat s}~
	 \bar {\it u}(p_3)(-ig_sT^A \gamma^{\mu}){\it v}(p_4),\\
\delta {\it M}& = &\bar {\it v}(p_2)(-ig_sT^A \gamma^{\nu}){\it u}(p_1)
	~\frac {-i g_{\mu\nu}} {\hat s}~
	 \bar {\it u}(p_3)\delta \Lambda^{\mu} {\it v}(p_4).
\end{eqnarray}
Here, $p_1, p_2$ denote the momenta of the incoming partons, and $p_3,p_4$ 
are used for outgoing t quark and its antiparticle. $\hat s$ is center-of-mass
energy of parton level process. $\delta \Lambda^{\mu}$ stand for the genuine 
supersymmetric electroweak correction to the vertex of top-quark color current,
which are given by
\begin{eqnarray}
\delta \Lambda^{\mu}&=&-ig_sT^A \frac {g^2 m_t^2}{32\pi^2 m_W^2 \sin^2 \beta}
\left [ \gamma^{\mu}F_1+\gamma^{\mu}\gamma_5 F_2\right.\nonumber\\
& & \left.+ k^{\mu} F_3  + k^{\mu} \gamma_5 F_4
+ i k_{\nu} \sigma^{\mu\nu} F_5
+ i k_{\nu} \sigma^{\mu\nu} \gamma_5 F_6 \right ],
\end{eqnarray}
where $\sigma^{\mu\nu}=\frac{i}{2} [\gamma^{\mu},\gamma^{\nu}]$ and the form 
factor $F_i$ are obtained by
\begin{equation}
F_i=F_i^c+F_i^n,
\end{equation}
where $F_i^c$ and $F_i^n$ arise from chargino and neutralino digrams, 
respectively. $F_i^c$ are given as
\begin{eqnarray}
F_1^c&=&\sum_{j=1,2}V_{j2}V_{j2}^* \left [ c_{24}+m_t^2(c_{11}+c_{21})
+(\frac{1}{2}B_1+m_t^2 B'_1)(m_t,\tilde M_j,m_{\tilde b})\right ]\\
F_2^c&=&\sum_{j=1,2}V_{j2}V_{j2}^* \left [ c_{24}
      +\frac{1}{2}B_1(m_t,\tilde M_j,m_{\tilde b}) \right ],\\
F_3^c&=&\frac{1}{2}m_t \sum_{j=1,2}V_{j2}V_{j2}^* 
		\left ( c_{21}-2c_{23} \right ),\\
F_4^c&=&\frac{1}{2}m_t \sum_{j=1,2}V_{j2}V_{j2}^*
	\left ( c_{21}+4c_{22}-4c_{23} \right ),\\
F_5^c&=&-\frac{1}{2}m_t \sum_{j=1,2}V_{j2}V_{j2}^*
	\left ( c_{11}+c_{21} \right ),\\
F_6^c&=&-\frac{1}{2}m_t \sum_{j=1,2}V_{j2}V_{j2}^*
	\left ( c_{11}+c_{21}-2c_{12}-2c_{23} \right ),\\
\end{eqnarray}
where the functions $c_{ij}(-p_3,p_3+p_4,\tilde M_j, m_{\tilde b},
m_{\tilde b})$ and $B_1$ are the 3-point and 2-point Feynman integrals [14].
The chargino masses $\tilde M_j$  and matrix elements $V_{ij}$ depend on 
parameters $M,\mu,\tan\beta$, whose expressions can be found in Ref.[12]. 
$B'_{0,1}$ are defined by 
\begin{equation}
B'_{0,1}(m,m_1,m_2)=
\frac{\partial B_{0,1}(p,m_1,m_2)}{\partial p^2}
				\vert_{p^2=m^2},
\end{equation}
$F_i^n$ are obtained by
\begin{eqnarray}
F_i^n&=&F_i^{\tilde t_1}+F_i^{\tilde t_2}+F_i^s{\rm~~(for~i=1,2})\\
F_i^n&=&F_i^{\tilde t_1}+F_i^{\tilde t_2}{\rm~~(for~i=3,4,5,6})
\end{eqnarray}
$F_1^s$ and $F_2^s$ are given by  
\begin{eqnarray}
F_1^s&=&\sum_{j=1}^4 \left \{ \frac{1}{2}N_{j4}N_{j4}^* \left [
B_1(m_t,\tilde M_{0j},m_{\tilde t_1})+B_1(m_t,\tilde M_{0j},m_{\tilde t_2})
\right ] \right. \nonumber\\
& & +m_t^2 N_{j4}N_{j4}^* \left [
B'_1(m_t,\tilde M_{0j},m_{\tilde t_1})+B'_1(m_t,\tilde M_{0j},m_{\tilde t_2})
\right ]  \nonumber\\
& & \left. +m_t \tilde M_{0j} N_{j4}N_{j4} \sin(2\theta) \left [
B'_0(m_t,\tilde M_{0j},m_{\tilde t_2})-B'_0(m_t,\tilde M_{0j},m_{\tilde t_1})
\right ] \right \} \\
F_2^s&=&\sum_{j=1}^4 \frac{1}{2}N_{j4}N_{j4}^* \cos(2\theta)
\left [ B_1(m_t,\tilde M_{0j},m_{\tilde t_1})
 -B_1(m_t,\tilde M_{0j},m_{\tilde t_2}) \right ] 
\end{eqnarray}
$F_i^{\tilde t_1}$ are given by
\begin{eqnarray}
F_1^{\tilde t_1}&=&\sum_{j=1}^4 \left\{ N_{j4}N_{j4}^* \left [ c_{24}
    		 +m_t^2(c_{11}+c_{21})\right ]
    -\sin(2\theta) N_{j4}N_{j4} m_t \tilde M_{0j}(c_0+c_{11})\right \}\\
F_2^{\tilde t_1}&=&\sum_{j=1}^4 N_{j4}N_{j4}^* c_{24}\cos(2\theta)\\
F_3^{\tilde t_1}&=&\sum_{j=1}^4 \left [ 
\frac{1}{2} m_t N_{j4}N_{j4}^* ( c_{21}-2c_{23} )
+\frac{1}{2}\sin(2\theta) N_{j4}N_{j4} \tilde M_{0j}(2c_{12}-c_{11})\right ]\\
F_4^{\tilde t_1}&=&\frac{1}{2} \cos(2\theta) m_t\sum_{j=1}^4 
        N_{j4}N_{j4}^* ( c_{21}+4c_{22}-4c_{23} )\\
F_5^{\tilde t_1}&=&\sum_{j=1}^4 \left [ 
-\frac{1}{2} m_t N_{j4}N_{j4}^* ( c_{11}+c_{21} )
+\frac{1}{2}\sin(2\theta) N_{j4}N_{j4} \tilde M_{0j}(c_0+c_{11})\right ]\\
F_6^{\tilde t_1}&=&-\frac{1}{2} \cos(2\theta) m_t\sum_{j=1}^4 
        N_{j4}N_{j4}^* ( c_{11}-2c_{12}+c_{21}-2c_{23} )
\end{eqnarray}
where $c_{ij}(-p_3,p_3+p_4,\tilde M_{0j}, m_{\tilde t_1},m_{\tilde t_1})$
are the 3-point Feynman integrals[14]. The neutralino masses $\tilde M_{0j}$ 
and matrix elements $N_{ij}$ are obtained by diagolis the matrix $Y$ [12].  
Giving the values for the parameters $M, M^{\prime},\mu,\tan\beta$,
the matrix $N$ and $N_D$ can be obtained numerically. Here, the parameters 
$M, M^{\prime}$ are the masses of gauginos corresponding to $SU(2)$ and $U(1)$,
respectively. With the grand unification assumption, i.e. $SU(2)\times U(1)$
is embedded in a grand unified theory, we have the relation
$M^{\prime}=\frac{5}{3}\frac{g'^2}{g^2} M$. $F_i^{\tilde t_2}$ are given by
\begin{equation}
F_i^{\tilde t_2}=F_i^{\tilde t_1}\left\vert_{
 	\sin(2\theta)\rightarrow -\sin(2\theta),
	\cos(2\theta)\rightarrow -\cos(2\theta),
	m_{\tilde t_1}\rightarrow m_{\tilde t_2}} \right.
\end{equation}

The renormalized cross-section for parton level process
$q\bar q \rightarrow t \bar t$ are given by 
\begin{equation}
\hat \sigma(\hat s)=\hat \sigma^0+\Delta \hat \sigma ,
\end{equation}
with
\begin{eqnarray}
\hat \sigma^0&=&\frac{8\pi \alpha_s^2}{27\hat s^2}\beta_t (\hat s+2m_t^2),\\
\Delta \hat \sigma&=&\frac{8\pi \alpha_s^2}{9\hat s^3}\beta_t
	\frac {g^2 m_t^2}{32\pi^2 m_W^2 \sin^2 \beta}
        \left [ \frac{2}{3}F_1\hat s (\hat s+2m_t^2)
	+2F_5 m_t \hat s^2  \right ],
\end{eqnarray}
where $\beta_t=\sqrt {1-4m_t^2/\hat s}$.

The hadronic cross section is obtained by convoluting the subprocess cross
section $\hat \sigma_{ij}$ of partons $i,j$ with parton distribution
functions $f^A_i(x_1,Q),f^B_j(x_2,Q)$, which is given by
\begin{equation}
\sigma(S)=\sum_{i,j}\int^1_{\tau_0} \frac{d\tau}{\tau} \left (\frac{1}{S}
	\frac{dL_{ij}}{d\tau}\right ) (\hat s \hat \sigma_{ij}),
\end{equation}
with
\begin{equation}
\frac{dL_{ij}}{d\tau}=\int^1_{\tau} \frac{dx_1}{x_1}
			\left [ f^A_i(x_1,Q)f^B_j(\tau/x_1,Q)
			+(A \leftrightarrow B  )\right ]
\end{equation}
In the above the sum runs over all incoming partons carrying a fraction of the
proton and antiproton momenta $(p_{1,2}=x_{1,2} P_{1,2})$, $\sqrt S=1.8$ TeV
is the center-of-mass energy of Tevatron, $\tau=x_1 x_2 $ and 
$\tau_0=4m_t^2/S$. As in Ref.[3], we do not distinguish the factorisation 
scale $Q$ and the renormalisation scale $\mu$ and take both as the top quark 
mass. In order to compare our results with the Yukawa corrections in Ref.[6],
we use the same parton distribution function as in ref.[6], i.e. the 
Morfin-Tung leading-order parton distribution function [15].

\begin{center}{\large 4. Numerical examples and discussion} \end{center}

 In the numerical examples presented in Figs.1-3, we fixed $M=200 GeV, 
\mu=-100 GeV$ and used the relation $M^{\prime}=\frac{5}{3}\frac{g'^2}{g^2} M$
to fix $M'$. Also we assumed $M_{\tilde t_R}=M_{\tilde t_L}=M_{\tilde b_L}$ 
which depend on sbottom mass $m_{\tilde b} \equiv m_{\tilde b_1}$ as in Eq.(6).
For $\tan\beta$ and the mixing parameter $M_{LR}$, we restrict them  to the 
range  $\tan\beta\leq 0.25 [6]$, $M_{LR}\leq 3m_{\tilde b_1}[16]$. 
Other input parameters are $m_Z=91.188 GeV,  ~\alpha_{em}=1/128.8 $, 
and $G_F=1.166372\times 10^{-5}(GeV)^{-2}$. $m_W$ is determined through $[17]$
\begin{equation}
m_W^2(1-\frac{m_W^2}{M_Z^2})=\frac{\pi\alpha}{\sqrt 2 G_F}\frac{1}{1-\Delta r},
\end{equation}
where, to order $O(\alpha m_t^2/m_W^2)$, $\Delta r$ is given by $[18]$
\begin{equation}
\Delta r\sim -\frac{\alpha N_Cc_W^2m_t^2}{16\pi^2s_W^4m_W^2}.
\end{equation}

The relative correction to the hadronic cross section as a function of sbottom
mass is presented in Fig.1 for $\tan\beta=0.25$ and  $\tan\beta=1$, 
respectively. Since the correction is proportional to $1/\sin^2\beta$, the 
size of correction for $\tan\beta=0.25$ is much larger than the corresponding
size for $\tan\beta=1$. In the range $m_{\tilde b}<150$ GeV, the correction 
is very sensitive to sbottom mass. The correction can be either negative or 
positive, depending on sbottom mass.  For $m_{\tilde b}>200$ GeV, the 
correction drops to about zero size, showing the  decoupling behaviour of MSSM.
Each plot in Fig.1 has a sharp dip, which occurs at the threshold point 
$m_t=m_{\tilde b}+\tilde M_j$. The chargino masses $\tilde M_{1,2}=(230, 100) 
GeV $ for  $\tan\beta=0.25$ and $\tilde M_{1,2}=(220, 120) GeV $ for  
$\tan\beta=1$, thus the threshold point locates at $m_{\tilde b}=76$ GeV 
for $\tan\beta=0.25$ and $m_{\tilde b}=56$ GeV for $\tan\beta=1$.  

Fig.2 show the dependence of the relative correction to the hadronic cross 
section on the value of $\tan\beta$ for sbottom mass $m_{\tilde b}=100 $ GeV
and 150 GeV, respectively. The correction is very sensitive to  $\tan\beta$
in the range $\tan\beta<1$. When $\tan\beta<<1$, the correction size  gets 
very large since it is proportional to $1/\sin^2\beta$.  
 
Fig.3 is the plot of the relative correction to the hadronic cross 
section versus the top-squark mixing parameter $M_{LR}$.
The corresponding neutralino masses in this figure are (122, 115, 77, 229) GeV
and chargino masses are $(230, 100)$ GeV.
 The starting point $M_{LR}=0$ correspondes to no-mixing case, at which
top-squark masses $m_{\tilde t_{1,2}}=m_{\tilde t_{L,R}}=(213, 216)$ GeV
and the mixing angel $\theta=0$. As  $M_{LR}$ increases, the mass splitting 
between two top-squarks increases. At  $M_{LR}=200$ GeV, 
top-squark masses $m_{\tilde t_{1,2}}=(103, 285)$ GeV
and the mixing angel $\theta=0.775$. The two sharp dips in the plot 
correspond to two threshold points at about $M_{LR}=200$ GeV and 230 GeV, 
at which $m_t=m_{\tilde t_1}+\tilde M_{0j}$.

So, from Figs.1-3 we found that only for $\tan\beta<1$ and $m_{\tilde b}<150$
GeV the correction size may exceed 20\%. For $\tan\beta\geq 1$ or 
$m_{\tilde b}>150$ the correction size can only reach a few percent.
In Ref.[6], the Yukawa correction from the SUSY Higgs sector to
the hadronic cross section was found to be very small for $\tan\beta=1$,
on the order of a percent, and only for minimum value $\tan\beta=0.25$ 
the correction can be above 10\% but never exceed 20\%. Therefore,  
the genuine supersymmetric electroweak corrections are comparable to
the Yukawa correction from the SUSY Higgs sector.

 In conclusion, we presented the anlytical expression for the genuine 
supersymmetric electroweak corrections of order $\alpha m_t^2/m_W^2$ to 
top quark pair production at the Tevatron with the consideration of
top-squark mixing. Numerical examples showed that only for $\tan\beta<1$ 
and $m_{\tilde b}<150$ GeV the correction can exceed 20\%.
In the most favorable case, these supersymmetric corrections, combined with
Yukawa corrections of the Higgs sector and also SUSY QCD correction,  
are potentially observable at Tevatron and could be used to place 
restrictions on MSSM.

\begin{center} {\bf ACKNOWLEDGMENTS } \end{center}  

  We thank  Jorge L. Lopez, Raghavan Rangarajan and Jaewan Kim for 
comparing their numerical results of neutralino masses and Feynman
integrals $B_{0,1}, c_0, c_{ij}$ with ours.

  This work was supported in part by the Foundation for Outstanding 
Young Scholars of Henan Province.

\vspace{1.in}
{\LARGE References}
\vspace{0.3in}
\begin{itemize}
\begin{description}
\item[{\rm [1]}] CDF Collaboration,  Phys.Rev.Lett. {\bf 74}, 2626(1995);\\
	        D0 Collaboration,  Phys.Rev.Lett. {\bf74}, 2632(1995).
\item[{\rm [2]}] P.Nason, S.Dawson and R.K.Ellis, Nucl.Phys. {\bf B303},
		609 (1988);\\
		G.Altarelli et al., Nucl.Phys. {\bf B308}, 724 (1988);\\
		W.Beenaker et al.,  Phys.Rev.D {\bf 40}, 54 (1989).
\item[{\rm [3]}] E.Laenen, J.Smith and W.L.van Neerven, Phys.Lett.
		{\bf B321},254 (1994).
\item[{\rm [4]}] E.L.Berger and H.Contopanagos, Argonne ANL-HEP-PR-95-31.
\item[{\rm [5]}] W. Beenakker et al., Nucl.Phys. {\bf B411}, 343 (1994).
\item[{\rm [6]} ] A Stange and S.Willenbrock, Phys.Rev. D{\bf 48}, 2054 (1993).
\item[{\rm [7]} ] C.S.Li, B.Q.Hu, J.M.Yang and C.G.Hu, Phys.Rev. D{\bf 52}, 5014(1995).
\item[{\rm [8]} ] J.M.Yang and C.S.Li, Phys.Rev. D{\bf 52}, 1541 (1995).
\item[{\rm [9]} ] J.M.Yang and C.S.Li, Erratum of Ref.[8], to appear in Phys.Rev. D.
\item[{\rm [10]}] J.Ellis and S.Rudaz, Phys.Lett.{\bf B128}, 248 (1983)\\
	A.Bouquet, J.Kaplan and C.Savoy, Nucl.Phys.{\bf B262}, 299 (1985). 
\item[{\rm[11]}] F.Berends, J.Tausk and W.Giele, Phys.Rev.D{\bf 47}, 2746(1993).
\item[{\rm[12]}] H. E. Haber and C. L. Kane, Phys. Rep. {\bf 117}, 75 (1985);\\
    	       J. F. Gunion and H. E. Haber, Nucl. Phys. {\bf B272}, 1 (1986).	
\item[{\rm[13]}] K.I.Aoki et al., Prog.Theor.Phys.Suppl. {\bf 73}, 1 (1982);\\
                M.Bohm, W.Hollik, H.Spiesbergerm, Fortschr. Phys. {\bf 34}, 687 (1986).
\item[{\rm[14]}]  A. Axelrod, Nucl. Phys. {\bf B209}, 349 (1982);\\
                  G. Passarino and M. Veltman, Nucl. Phys. {\bf B160}, 151 (1979);
\item[{\rm[15]}] J.Morfin and W.K.Tung,  Z.Phys. {\bf C52}, 13 (1991).
\item[{\rm[16]}] D.Garcia, R.A.Jimenez, J.Sola and W.Hollik, Nucl. Phys.
                    {\bf B427}, 53 (1994). 
\item[{\rm[17]}]  A. Sirlin, Phys. Rev. D{\bf 22}, 971 (1980);\\
            W. J. Marciano and A. Sirlin, Phys. Rev. D{\bf 22}, 2695 (1980);
            (E) D{\bf 31}, 213 (1985);\\
            A. Sirlin and W. J. Marciano, Nucl. Phys. {\bf B189}, 442 (1981);\\
            M. B\"ohm, W. Hollik and H. Spiesberger, Fortschr. Phys. {\bf 34},
            687 (1986).
\item[{\rm[18]}] W. J. Marciano and Z. Parsa, Annu. Rev. Nucl. Sci.{\bf 36}, 171 (1986);\\
                W. Hollik, Report No. CERN 5661-90 (1990) (unpublished).
\end{description}
\end{itemize}
\vfil
\eject

\begin{center} {\bf Figure Captions} \end{center}

Fig.1 The plot of relative correction to the hadronic cross section 
versus sbottom mass, where $M_{LR}=1.5m_{\tilde b}$.

Fig.2 The plot of relative correction to the hadronic cross 
section versus $\tan\beta$, where $M_{LR}=1.5m_{\tilde b}$.
 
Fig.3 The plot of the relative correction to the hadronic cross 
section versus the top-squark mixing parameter $M_{LR}$.

\end{document}